\documentclass{evn2024}
\usepackage{natbib}
\usepackage{graphicx}
\usepackage{amsmath}
\usepackage{url}
\usepackage{epstopdf}
\begin{document}
   \title{High-resolution radio observations of TeV candidate sources}

  \author{J. K\H{o}m\'{\i}ves
          \inst{1},
          K. \'E. Gab\'anyi\inst{1,3,4,5},
          S. Frey\inst{2,4,5},
          Z. Paragi\inst{6},
          T. An\inst{7}
          \and
          E. Kun\inst{4,5,8}
          }
   \institute{Department of Astronomy, Institute of Physics and Astronomy, ELTE E\"otv\"os Lor\'and University, P\'azm\'any P\'eter s\'et\'any 1/A, H-1117 Budapest, Hungary
         \and 
            Institute of Physics and Astronomy, ELTE E\"otv\"os Lor\'and University, P\'azm\'any P\'eter s\'et\'any 1/A, H-1117 Budapest, Hungary
         \and
             HUN-REN--ELTE Extragalactic Astrophysics Research Group, ELTE E\"otv\"os Lor\'and University, P\'azm\'any P\'eter s\'et\'any 1/A, H-1117 Budapest, Hungary
        \and
             Konkoly Observatory, HUN-REN Research Centre for Astronomy and Earth Sciences, Konkoly Thege Mikl\'os \'ut 15-17, 1121 Budapest, Hungary      
        \and
             CSFK, MTA Centre of Excellence, Konkoly Thege Mikl\'os \'ut 15-17, 1121 Budapest, Hungary
        \and
             Joint Institute for VLBI ERIC, Oude Hoogeveensedijk 4, 7991 PD Dwingeloo, The Netherlands
        \and
             Shanghai Astronomical Observatory, 80 Nandan Road, 200030 Shanghai, People’s Republic of China 
        \and
             Faculty for Physics \& Astronomy, Ruhr University Bochum, Germany\\
            }

   \abstract
   {Radio-loud active galactic nuclei (AGN) with their jets pointed close to our line of sight constitute the majority of extragalactic $\gamma$-ray sources and significantly contribute to the radiation observed in the even higher energy regime. The upcoming Cherenkov Telescope Array (CTA) is expected to detect fainter TeV objects, leading to an anticipated increase in the proportion of non-blazar extragalactic high-energy sources.
   Here we present the results of our dual-frequency ($1.7$ and $5$~GHz) European VLBI Network (EVN) and enhanced Multi Element Remotely Linked Interferometer Network (e-MERLIN) observations of two faint radio sources from the list of TeV candidate sources. They do not show signs of nuclear activity in their optical spectra, but they were hypothesized to contain faint AGN that is outshone by the host galaxy. 
   We used the mas-scale resolution radio data to try to pinpoint the location of the compact radio emitting feature, determine its spectral index, radio power, brightness temperature and radio--X-ray luminosity ratio and thus identify the origin of the radio emission.
   Our results suggest that both optically passive-looking galaxies host faint compact radio-emitting AGN with steep spectra. }

\titlerunning{High-resolution radio observations of TeV candidate sources}
\authorrunning{J. K\H{o}m\'{\i}ves et al.}   
\maketitle

%
%________________________________________________________________

\section{Introduction}

Blazars, a subclass of radio-loud active galactic nuclei (AGN), are characterized by their jets aligning closely with the observer’s line of sight. They are categorized into BL Lacertae objects (BL Lacs), which exhibit non-existent or very weak emission lines, and Flat-Spectrum Radio Quasars (FSRQs), which show strong emission lines in their optical spectra. The spectral energy distributions (SEDs) of blazars are primarily influenced by non-thermal emission processes: synchrotron radiation and inverse-Compton emission. These processes manifest as two broad peaks in the SED, with synchrotron emission spanning from radio to optical or X-rays, and inverse-Compton emission extending from X-rays to GeV or TeV $\gamma$-rays. Blazars also constitute the largest fraction of the extragalactic $\gamma$-ray-emitter objects and expected to be the most common and most numerous class of extragalactic objects detectable at TeV energies with the future Cherenkov Telescope Array \cite[CTA,][]{CTA2019}.

\cite{balmaverde} compiled a list of $87$ radio- and X-ray-detected objects based on the NRAO VLA Sky Survey \citep[NVSS, ][]{condon1998} and the ROentgen SATellite Positional Sensitive Proportional Counter (ROSAT PSPC) catalogues \citep{rosat,wgacat} to provide a list of good TeV-emitting candidate objects. Apart from the radio and X-ray detections, they restricted themselves to objects with high X-ray-to-radio flux density ratios, since most of the known TeV-detected blazars exhibit similar characteristics \citep{balmaverde}. Using archival and their own optical spectroscopic data, they excluded the objects with strong emission lines. They classified remaining $46$ objects (TeV-emitting Radio-Emitting X-ray sources, Te-REXes) into two groups, BL Lacs and Passive Elliptical Galaxies (PEGs) depending on whether the properties of the optical spectra around the CaII absorption spectral line indicate the presence of non-thermal emission originating from an AGN \citep{Stocke1991}, or not. \cite{balmaverde} argue that the PEGs could be part of the ``missing'' population of beamed-up but inherently low radio luminosity AGN \citep{Marcha2013}.

We observed at mas-scale resolution two PEGs, 1$\mathrm{REX\,J}$151924$+$2053.7 and 1$\mathrm{REX\,J}$183200$+$5202.2 (hereafter J1519 and J1832), expected to be bright enough at TeV ranges to be detectable with the CTA, to identify the origin of their radio emission.

In the following, we assume a flat $\Lambda$ Cold Dark Matter cosmological model with Hubble constant $H_0=70$~km\,s$^{-1}$\,Mpc$^{-1}$, matter density parameter $\Omega_\mathrm{m}= 0.27$, and vacuum energy density parameter $\Omega_\mathrm{vac} = 0.73$. At the redshifts of J1519 \citep[$z=0.041$, ][]{2015ApJS..219...12A}, and J1832 \citep[$z=0.046$, ][]{balmaverde}, the angular scales are $0.80$\,pc\,mas$^{-1}$ and $0.91$\,pc\,mas$^{-1}$, respectively \citep{wright}.

\section{Observations and data reduction}

\begin{table*}
\caption[]{Details of the EVN observations. Note: telescope codes: Jodrell Bank (Jb, UK), Westerbork (Wb, Netherlands), Effelsberg (Ef, Germany), Medicina (Mc, Italy), Noto (Nt, Italy), Onsala (O8, Sweden), Tianma (T6, China), Urumqi (Ur, China), Toru\'n (Tr, Poland), Yebes (Ys, Spain), Hartebeesthoek (Hh, South Africa), Cambridge (Cm, UK), Darnhall (Da, UK), Knockin (Kn, UK), Pickmere (Pi, UK), Defford (De, UK)}
\label{table:obs}
\centering
    \begin{tabular}{c c c}
    \hline
    \vtop{\hbox{\strut $\textbf{Frequency}$}\hbox{\strut [GHz]}} & \vtop{\hbox{\strut $\textbf{Observing date}$}\hbox{\strut [yyyy-mm-dd]}} & $\textbf{Participating EVN telescopes}$ \\
    \hline
    $1.7$ & 2022-03-08 & Jb, Wb, Ef, Mc, Nt, O8, T6, Ur, Tr, Hh, Cm, Da, Kn, Pi, De \\
    $5$ & 2022-03-15 & Jb, Wb, Ef, Mc, Nt, O8, T6, Ur, Tr, Ys, Hh, Cm, Da, Kn, Pi, De \\
    \hline
    \end{tabular}
\end{table*}

\begin{table*}
\caption[]{Properties of the target sources: name, redshift, flux density from the NVSS and VLASS, the luminosity distance of the source and the name of the corresponding phase calibrator source. Coordinates are from NVSS.}
\label{table:targ}
\centering
    \begin{tabular}{c c c c c c}
    \hline
    $\textbf{Source ID}$ & $\mathbf{z}$ & $\mathbf{S_{1.4, NVSS}}$ [mJy] & $\mathbf{S_{3, VLASS}}$ [mJy] & $D_{\mathrm{L}}$ [Mpc] & $\textbf{Phase calibrator}$ \\
    \hline
    $\mathrm{J}$1519 & $0.041$ & $16.10 \pm 0.14$ & $7.12 \pm 0.12$ & $178.5$ & $\mathrm{J}$1516$+$1932 \\
    $\mathrm{J}$1832 & $0.046$ & $9.6 \pm 0.5$ & $9.4 \pm 0.11$ & $204.1$ & $\mathrm{J}$1816$+$5307 \\
    \hline
    \end{tabular}
\end{table*}

\begin{table*}
\caption[]{Fitted and calculated parameters of the sources: name, frequency ($\nu$), flux density ($S$), The FWHM sizes of major and minor axes of the Gaussian components: $W_1$ and $W_2$ (In the case of circular components, only $W_1$ is given.), EVN position of the brightest pixel (right ascension, declination), brightness temperature ($T_\mathrm{b}$) and radio power ($P$) at $1.7$\,GHz.}
\label{table:parameters}
\centering
    \begin{tabular}{c c c c c c c c c}
    \hline
    $\textbf{Source ID}$ & 
    \vtop{\hbox{\strut$\mathbf{\nu}$}\hbox{\strut[GHz]}} & 
    \vtop{\hbox{\strut$\mathbf{S}$}\hbox{\strut[mJy]}} & \vtop{\hbox{\strut$\mathbf{W_1}$}\hbox{\strut[mas]}} & \vtop{\hbox{\strut$\mathbf{W_2}$}\hbox{\strut[mas]}} &  \vtop{\hbox{\strut$\textbf{RA}$}\hbox{\strut[hh mm s.sss]}} & \vtop{\hbox{\strut$\textbf{Dec}$}\hbox{\strut[$^{\circ}$ $^{\prime}$ $^{\prime\prime}$]}} & \vtop{\hbox{\strut$\mathbf{T_b}$}\hbox{\strut[$10^{7}$ K]}} & \vtop{\hbox{\strut$\mathbf{P}$}\hbox{\strut[$10^{21}$\,W\,Hz$^{-1}$]}} \\
    \hline
    $\mathrm{J}$1519 & 1.7 & $ 4.2 \pm 0.5 $ & $ 8.0 \pm 1.2 $ &  $ 1.9 \pm 0.3 $ & 15 19 24.7366 & 20 53 46.6662 &  $ 3.3 \pm 0.2 $ &  $ 13.0 \pm 1.0 $  \\
    & & $ 0.7 \pm 0.1 $ & $ 13.6 \pm 1.2 $ & - & - & - & - & - \\
    & 5 & $ 2.5 \pm 0.5 $ & $ 3.4 \pm 0.3 $ &- & 15 19 24.7368 & 20 53 46.6679 & $ 10.9 \pm 0.3 $ & $ 7.8 \pm 0.8$  \\
    & & $ 1.0 \pm 0.1 $ & $ 3.0 \pm 0.3 $ & - & - & - & - & -  \\
    \hline
    $\mathrm{J}$1832 & 1.7 & $ 7.2 \pm 0.4 $ & $ 6.8 \pm 0.4 $ & $ 1.4 \pm 0.4 $ & 18 32 00.6513 & 52 02 18.1792 &  $ 7.7 \pm 0.2 $ &  $ 29.0 \pm 3.0 $  \\
    & & $ 1.4 \pm 0.1 $ & $ 10.7 \pm 0.5 $ & - & - & - & - & -  \\
    & 5 & $ 9.0 \pm 1.1 $ & $ 3.2 \pm 0.5 $ &- & 18 32 00.6514 & 52 02 18.1780 & $ 5.1 \pm 0.3 $ & $ 36.0 \pm 4.0 $  \\
    \hline
    \end{tabular}
\end{table*}

\begin{figure}
   \centering
   \includegraphics[width=8.8cm]{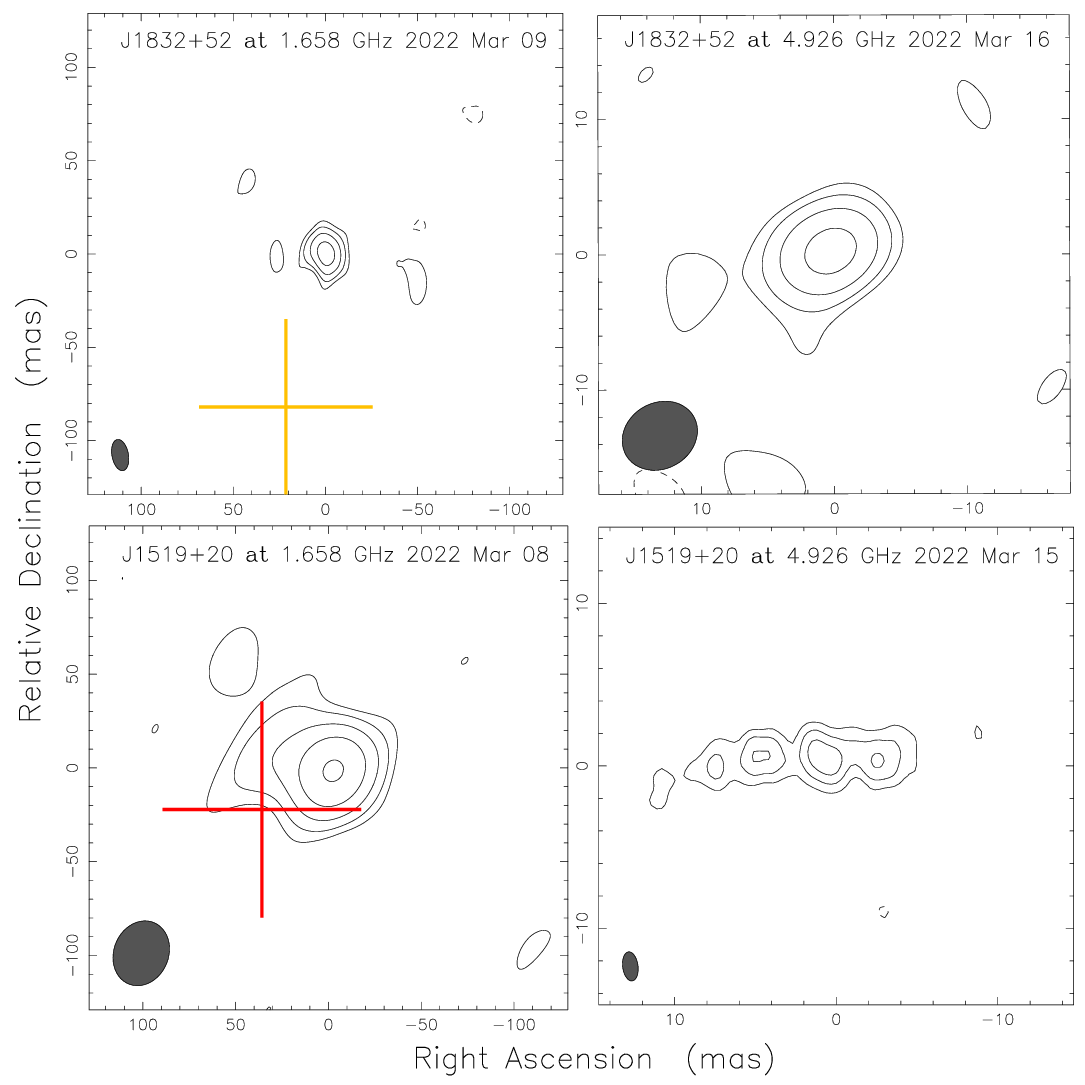}
   \vspace{0pt}
%   \special{psfile=JKomives_Fig1-3.ps hscale=55 vscale=55 hoffset=10 voffset=0}
   \caption{VLBI images of sources J1832 (top) and J1519 (bottom) at at $1.7$~GHz (left) and $5$~GHz (right). {Panel} (\textbf{a}): The peak intensity is $4.41$ mJy\,beam$^{-1}$. The lowest contours are at $\pm 0.31$\,mJy\,beam$^{-1}$. The restoring beam size is $16.9\mathrm{\,mas} \times 9.0\mathrm{\,mas}$. The position angle of the major axis is $\mathrm{PA}=11.1^{\circ}$. {Panel} (\textbf{b}): The peak intensity is $0.48$\,mJy\,beam$^{-1}$. The lowest contours are at $\pm 0.48$\,mJy\,beam$^{-1}$. The restoring beam size is $5.8\mathrm{\,mas} \times 4.9\mathrm{\,mas}$, $\mathrm{PA} = -62.5 ^{\circ}$. {Panel} (\textbf{c}): The peak intensity is $3.32$\,mJy\,beam$^{-1}$. The lowest contours are at $\pm 0.21$\,mJy\,beam$^{-1}$. The restoring beam size is $35.4 \mathrm{\,mas}\times 29.5\mathrm{\,mas}$, $\mathrm{PA} = -23.5 ^{\circ}$. {Panel} (\textbf{d}): The peak intensity is $1.34$\,mJy\,beam$^{-1}$. The lowest contours are at $ \pm  0.06$\,mJy\,beam$^{-1}$. The restoring beam size is $1.8 \mathrm{\,mas}\times 1.0\mathrm{\,mas}$, $\mathrm{PA} = -7.6 ^{\circ}$.\linebreak In each map, the lowest positive contour is drawn at the $3.5 \sigma$ image noise level, and subsequent contours increase by a factor of two. The restoring beams are shown in the lower left corners of each panel. Red cross denotes the position of the nearest optical galaxy to J1519 detected in the SDSS. Yellow cross indicates the the nearest Gaia DR3 \citep{Gaia} object’s position to J1832. The size of the crosses indicates the uncertainty of the optical positions.)
            \label{fig:cleans}
           }
\end{figure}

The two targets were observed with the European VLBI Network (EVN) and the enhanced Multi Element Remotely Linked Interferometer Network (e-MERLIN) networks under the project code EG110 in phase-referencing mode \citep{beasley}. The observations were performed in two frequency bands, at $5$\,GHz and at $1.7$\,GHz. There were $8$ intermediate frequency channels (IFs) at $5$\,GHz and $4$ at $1.7$\,GHz, with $32$\,MHz bandwidth each. %Sixty-four $500$\,kHz spectral channels were included in each of these subband. 
Thus the total bandwidths were $256$\,MHz for $5$\,GHz and $128$\,MHz for the $1.7$\,GHz observations.
Table \ref{table:obs} lists the details of the observations, including frequency, date, and participating telescopes. Each target was observed for $\approx 3$\,h, alternating between the corresponding phase-reference source and the target in 5-min cycles, with  $3.5$~min spent on the target every cycle. The two phase-reference sources were NVSS\,J151656$+$193213 and NVSS\,J181657$+$530744 (hereafter J1516$+$1932 and J1816$+$5307). Additional $3$ calibrator sources were observed as fringe finders. Phase-reference calibrators were chosen from the Astrogeo website\footnote[1]{\url{http://astrogeo.org}, maintained by L. Petrov} \citep{Petrov2024}.

For data reduction, we used the National Radio Astronomy Observatory (NRAO) Astronomical Image Processing System \citep[AIPS,][]{greisen} software package and we followed the standard procedures of the EVN data reduction guide\footnote[2]{\url{https://www.evlbi.org/evn-data-reduction-guide}}. Details of the target sources can be seen in Table \ref{table:targ}. 

\section{Results}

We produced clean images of both targets at both frequencies. The intensity maps of the targets are shown in Fig.~\ref{fig:cleans}. Most of our maps show no extended structures with only one significant radio feature detected. In case of the $5$\,GHz VLBI map of J1519$+$2053, we can see that the structure is elongated in the east--west direction and extended to about $20$\,mas.

We determined the VLBI positions of the brightness peaks using the AIPS task $\textsc{MAXFIT}$ (Table \ref{table:targ}).
When calculating the positional uncertainty, we took into account the known errors of the positions of phase-reference sources ($0.10$\,mas for J1516$+$1932 and $0.14$\,mas for J1816$+$5307), and the frequency-dependent uncertainty coming from the angular separations between the targets and their respective phase-reference calibrators. This latter was the dominant component with $\sim$ $0.5$\,mas in the case of J1519 and $\sim$ $3$\,mas in the case of J1832 at $5$~GHz \citep{Chatterjee}. In the end, the total positional uncertainties of the two targets are $\sim0.6$\,mas and $\sim3$\,mas respectively.

In order to quantify the flux density and the size of the source, we fitted the visibility data with two-dimensional Gaussian model components. At $5$~GHz, two circular model components were needed to adequately fit the visibility data of J1519 while one for J1832. At $1.7$~GHz, the visibility data of both targets could be well described with one elliptical, for the brightest features, and one circular Gaussian components. The parameters of the model fit are given in Table \ref{table:parameters}. The lack of self-calibration can result in coherence loss causing the flux density to be less than it would be with the phase correction. We accounted for this coherence loss with a typical amount of $20\%$ recovered flux density \citep{Gabanyi,Mosoni}. For the error calculations of the model parameters, we used a modified version of the formulae described in \citep{Image_Analysis}, taking into account the image artifacts arising from the sparse $(u,v)$ coverage of VLBI observations \citep{Emma2014, Schinzel}. 

Using the dual-frequency observations, we calculated the spectral indices of the brightest features ($\alpha$), with the convention $S_\mathrm{core}\propto \nu^\alpha$, where $\nu$ is the frequency. These values ($ -0.45 \pm 0.13 $ for J1516$+$1932 and $ -0.21 \pm 0.07 $ for J1816$+$5307) are indicating ﬂat radio spectra usually attributed to the base of the jet in radio-emitting AGN \citep[e.g.,][]{hovatta}. 

From the parameters of the model fitting (Table \ref{table:parameters}), we calculated the brightness temperatures and radio powers of the target sources. The brightness temperature $T_\mathrm{b, VLBI}$ can be calculated for the fitted Gaussian core component knowing the redshift of the object \citep[e.g.,][]{condon1982}:
\begin{equation}
    T_{\mathrm{b,VLBI}}=1.22 \cdot 10^{12} (1 + z) \frac{S_\mathrm{core}}{\nu^2 W_1 W_2} \text{\, [K],}
\end{equation}
where $S_\textrm{core}$ is the flux density expressed in Jy, and $\nu$ is the observing frequency in GHz. The component major and minor axes (full width at half maximum, FWHM) $W_1$ and $W_2$ are given in mas. We found that both cores have brightness temperatures higher than the limit for galaxies without active nuclei at $1.7$~GHz frequency \citep[$T_\mathrm{b} \approx 10^5$\,K,][]{condon1992}.

We estimated the radio powers of the cores of our targets using the formula:
\begin{equation}
    P = 4\pi D_{\mathrm{L}}^2 S_\mathrm{core} (1 + z)^{-\alpha - 1},
\end{equation}
where $D_{\mathrm{L}}$ denotes the luminosity distance. The $1.7$~GHz radio powers of the two sources are in the order of $10^{22}$\,W\,Hz$^{-1}$, thus they are an order of magnitude higher than the limiting value for non-thermal radio emission originating from starburst-related activity \citep[$\sim 2 \cdot 10^{21}$\,W\,Hz$^{-1}$,][]{kewley,middelberg}. 

We compared the flux density of our VLBI measurements with larger-scale radio observations. Table \ref{table:targ} shows the flux densities measured with the NVSS and recently with the Very Large Array Sky Survey \citep[VLASS,][]{vlass}. Our $1.7$~GHz VLBI observations resolved out significant amount of flux density, $\sim 11.2 $\,mJy for J1519, while for J1832, the missing flux density is lower, $\sim 1.0 $\,mJy -- assuming there was no flux density variability in the objects. If the cause of the missing flux density is the resolution difference, then it originates from structures with scales larger than $\sim 750$\,mas, which was the largest recoverable size in our observations at $1.7$~GHz defined by the smallest baseline length, between Darnhall and Knockin ($\sim 50 $\,km).

\section{Discussion and summary}

We compared the positions of the peak of the $5$~GHz radio images with the optical positions of the corresponding sources. 
In the case of J1832, the radio and optical positions, the latter taken from the Gaia DR3 \citep{Gaia}, have large difference (see Fig. \ref{fig:cleans}). However, the Gaia value has an astrometric excess noise of $46$\,mas with a significance of $116$, indicating that the observation disagrees with the best-fitting standard astrometric model \citep{Lindegren}. There is no detection at the position of this source in the Sloan Digital Sky Survey (SDSS) \citep{SDSS}. Thus, given the quality of the optical data, we cannot draw any conclusions about the offset of our target from the centre of its host galaxy.
In the case of J1519, there is no optical counterpart detected in Gaia DR3 at the radio position, but there is an SDSS-detected galaxy coinciding with J1519 within errors (see Fig. \ref{fig:cleans}). We note that this galaxy seems slightly elongated in the east-west direction, similarly to the structure of our $5$~GHz radio map.

The relatively high brightness temperatures, and flat radio spectra of the targets detected at mas scale resolutions strongly indicate AGN origin of the radio emission. Even though, due to the lack of precise optical positions, we cannot ascertain whether the radio emission in J1832 indeed originates at the centre of its host galaxy.

According to \cite{LaorBehar}, the radio emission in weak radio-emitting AGN can originate from the corona. They show that the empirically derived relationship between the radio ($L_\mathrm{R}$) and X-ray luminosity ($L_\mathrm{X}$) for coronally active stars, $L_\mathrm{R}/L_\mathrm{X} \approx 10^{-5}$, can apply for some radio quiet AGN. For our targets with the used X-ray flux densities given in the ROSAT catalog, measured in the energy range of $0.5-2$\,keV, and the radio flux densities at $5$~GHz, the luminosity ratio aligns with this $10^{-5}$ threshold. This suggests that the faint radio emissions of our PEG sources could potentially originate from the X-ray corona.

\begin{acknowledgements}

The European VLBI Network is a joint facility of independent European, African, Asian, and North American radio astronomy institutes. Scientific results from data presented in this publication are derived from the following EVN project code: EG110.
e-MERLIN is a National Facility operated by the University of Manchester at Jodrell Bank Observatory on behalf of STFC. On behalf of the ``Interferometric studies of radio-loud active galactic nuclei'' project, we are grateful for the possibility to use the HUN-REN Cloud \citep{H_der_2022} which helped us achieve the results published in this paper. This project was supported by the Hungarian National Research, Development and Innovation Office (NKFIH OTKA K134213) and by the HUN-REN. KÉG and SF also received funding from the NKFIH excellence grant TKP2021-NKTA-64. 

\end{acknowledgements}
\bibliographystyle{aa} 
\bibliography{references}
%\begin{thebibliography}{}
    
%    \bibitem[2020]{balmaverde} Balmaverde, B. et al. 2020, \mnras, 492, 3728-3741
      
%   \bibitem[1998]{condon1998} Condon, J. J. et al. 1998, AJ, 115, 1693

%   \bibitem[1996]{voges} Voges, W. et al. 1996, IAU Circ. 6420

%   \bibitem[2023]{gaia} Gaia Collaboration et al. 2023, A\&A, 674, A1

%   \bibitem[1992]{condon1992} Condon, J. J. et al. 1992, ARA\&A 30, 575
      
%   \bibitem[2000]{kewley} Kewley, L. J. et al. 2000, ApJ 530, 704
   
%   \bibitem[1987]{middleberg} Middelberg, E. et al. 2011, A\&A 526, A74
  
%   \bibitem[1992]{hovatta} Hovatta, T. et al. 2014, AJ 147, 143

%\end{thebibliography}

\end{document}